\documentclass{ifacconf}

\usepackage{graphicx}      
\usepackage{natbib}        
\usepackage{epstopdf}
\usepackage{amsmath}
\usepackage{amsfonts}
\usepackage{dsfont}
\usepackage{amssymb}
\usepackage{color}

\newtheorem{definition}{{Definition}} 
\newtheorem{remark}{{Remark}} 
\newtheorem{theorem}{Theorem}
\newtheorem{corollary}{Corollary}
\newtheorem{proposition}{Proposition}

\newtheorem{lemma}{Lemma}
\begin{document}
\begin{frontmatter}

\title{Passivity based distributed tracking  control of networked Euler-Lagrange systems}


\author[First,Second]{Rodolfo Reyes-B\'aez}
\author[First,Second]{Arjan van der Schaft} 
\author[First,Second]{Bayu Jayawardhana}

\address[First]{Jan C. Willems Center for Systems and Control, Faculty of Science and Engineering, University of Groningen, The Netherlands}

\address[Second]{Faculty of Science and Engineering, University of Groningen, P.O. Box 407, 9700 AK, Groningen, The Netherlands\\ (\{r.reyes-baez, a.j.van.der.schaft, jayawardhana\}@rug.nl)}


\begin{abstract}                
In this paper we present three distributed control laws for the coordination of networked Euler-Lagrange (EL) systems. We first reformulate the passivity-based control design method in \cite{Arcak} by considering that each edge is associated with an \emph{artificial spring system} instead of the usual diffusive coupling among the communicating agents. With this configuration, the networked EL system possesses a "symmetric" feedback structure which together with the strict passivity of both agents' and edges' dynamics lead to a strictly passive network dynamics. Subsequently we present the networked version of two different passivity-based tracking controllers 
that are particular cases of our method and the one in \cite{Arcak}. Numerical simulation is presented to show the performance of the proposed methods.

\end{abstract}


\end{frontmatter}

\section{Introduction}

The use of collaborative robots (which include mobile robots and UAVs) and of networked electro-mechanical systems are pervasive in various application domains, such as, smart factories, smart logistic systems, intelligent buildings and smart grids. For instance, the collaborative robots can be deployed to solve a variety of different tasks by autonomously coordinating their movements and actions among themselves. As another example, a network of machines in the shop floor of a smart factory can reconfigure themselves cooperatively and autonomously to produce a variety of different products. Against the backdrop, the distributed control  methods thereof have been an active area of research for the past decade, providing provably-correct control algorithms that can guarantee the completion of every given task by the group of robots or by the networked machines. These physical systems are typically belong to the class of Euler-Lagrange (EL) systems and the distributed tracking control method of networked EL systems is the focus of this paper. 

For a single EL system, passivity-based control (PBC) techniques have become the {\it de-facto} control design framework for a single EL system due to their natural  interpretation of the closed-loop systems, see, for example the books of  \cite{ortega2013passivity} and \cite{arjanl2}. In this framework, physical energy variables of the EL systems are explicitly used in the control design, the systems' interconnection is translated to energy exchange among different systems and the total energy dissipation is shaped in such a way that it fulfills the transient behaviour requirement and the state asymptotically converge to the minimum of the total energy. The term passivity-based control was introduced in \cite{ortega1988adaptive}. Recent results on PBC for a single EL system are the works of \cite{Lagrangians}, \cite{rodolfo2017}, \cite{rodolfo2018cdc} which incorporate recent advances in contraction theory and incremental stability, such as, \cite{jouffroy} and \cite{andrieu2016tac}. 

The passivity of EL mechanical systems is further exploited in the work of \citet{slotineli} where a so-called \emph{virtual system} is used as target closed-loop system in the control design procedure. Remarkably, such virtual system inherits the passivity of the actual one and it has been shown recently that this virtual system has \emph{contraction} properties in \cite{jouffroy}. This observation was recently exploited for \emph{differential passivity based control} design in \cite{Lagrangians}, \cite{rodolfo2017} and \cite{rodolfo2018cdc}. Recently, the passivity approach has been extended  to solve the distributed control problem for networked EL systems. 


The generalization of these PBC methods to the multi-agent setting has been well-studied in recent decade. The book of \cite{Arcak}, \cite{bai2011cooperative}, \cite{arjanl2} and the article by \cite{chopra2006passivity} provide a thorough exposition to the design of passivity-based distributed control methods where a number of coordination control problems can be solved through PBC approach, including, synchronization and formation control. For networked EL systems, some relevant works are the articles by \cite{marina2018}, \cite{nuno2013automatica}, \cite{nuno2013ejc} and \cite{Chung}.   


In the present paper, we reformulate the PBC design methodology in \cite{Arcak} where, roughly speaking, the coordination control problem is solved by interconnecting (strictly passive) systems attached to the nodes of a graph via diffusive coupling that preserves the  passivity of the network dynamics. As an alternative,  we attach strictly passive \emph{artificial spring systems} to each node and they are feedback interconnected to nodes dynamics. This results in dynamics protocols where the spring dynamics can be interpreted as a (nonlinear) integral action. Due to the strict passivity of the interconnected system, the asymptotic stability result can be established by using the total storage function as a \emph{strict} Lyapunov function. 
We also present two other distributed control methods which use networked virtual systems and can be seen as a particular case of our proposed method, as well as, the one in \cite{Arcak}.




The structure of the paper is as follows: In Section 2 we present the  preliminaries of graph theory and  mechanical systems in the EL framework that are necessary for the network dynamics modeling. Moreover, we recall two well-known passivity-based tracking controllers for a single mechanical system. In Section 3 we present the main contribution of our work, where instead of diffusive coupling we attach an artificial spring system to each node. The interconnection of the nodes and edges exhibits a symmetric structure that preserves passivity. Two other diffusive coupling protocols are presented in Section 4 which can be seen as particular cases of the proposed methodology exploiting the concept of virtual system for the network dynamics. Section 5 presents a simulation result that show the performance of our proposed approach in Section 3. 

\section{Networked Euler-Lagrange systems   preliminaries}\label{sII}
In this work we consider a network of  $N$  EL systems (agents) which interact among themselves for solving tracking control problem in a \emph{coordinated manner}. The  interaction  among  agents in the network is represented by the links of a {graph}.

\subsection{Graph theory tools }
The following  preliminaries of graph theory are taken from \cite{bollobas1998modern} and \cite{van2013port}.
\begin{definition}[Graph]
	A graph $\mathcal{G}$ is defined by a pair $(\mathcal{V},\mathcal{E})$ where $\mathcal{V}$ is a finite set of N vertices (also called nodes)  and $\mathcal{E}$ the finite set of $M$ edges (also called links). Furthermore, there is an injective mapping  from $\mathcal{E}$ to the set of unordered pairs $\mathcal{V}$, identifying edges with unordered pairs of vertices.
\end{definition}
The set $\mathcal{V}$ is called the vertex set of the graph $\mathcal{G}$, and the set $\mathcal{E}$ is called 
the edge set. The graph $(\mathcal{V},\mathcal{E})$ is said to be \emph{directed} if the edges are \emph{ordered} pairs $e_{ij} = (w_i,w_j )$. Then $w_i$ represents the tail vertex and $w_j$ the head vertex of $e_{ij}$. If $ e_{ij} = (w_i,w_j) \in \mathcal{E}$, then the vertices $w_i$ and $w_j$ are called \emph{adjacent} or \emph{neighboring} vertices; and $w_i$
and $w_j$ are \emph{incident} with the edge $e_{ij}$. Two edges are
\emph{adjacent} if they have exactly one common end-vertex. A
\emph{path} (weak path) of length $r$ in a directed graph is a
sequence $w_1,\dots,w_\iota$ of $\iota+1$ distinct vertices such that
for each $i\in\{1,\dots,\iota\}$, $(w_i,w_{i+1}) \in \mathcal{E}$ (respectively,
either $(w_i,w_{i+1}) \in \mathcal{E}$ or $(w_{i+1},w_{i}) \in \mathcal{E}$. A directed
graph is \emph{strongly connected} (weakly connected) if any two vertices can be joined by a path (respectively, weak path).

Given a graph $\mathcal{G}$, we define its \emph{vertex space} $\Lambda_0(\mathcal{G})$ as the vector space of all functions from $\mathcal{V}$ to some linear space $\mathcal{R}$. Furthermore, we define the \emph{edge space} $\Lambda_1(\mathcal{G})$ as the vector space of all functions from $\mathcal{E}$ to $\mathcal{R}$. The dual spaces of $\Lambda_0$ and $\Lambda_1$ will be denoted by $\Lambda^0$ and $\Lambda^1$, respectively. For a directed graph $\mathcal{G}$, the \emph{incidence operator}   is a linear map $B:\Lambda_1\rightarrow \Lambda_0$ with matrix representation $\hat{B}\otimes I$, where 
\begin{equation}
\hat{B}\otimes I:= \begin{bmatrix}
\hat{b}_{11}I&\dots& \hat{b}_{1M}I\\
\vdots & \ddots & \vdots\\
\hat{b}_{N1}I&\dots& \hat{b}_{NM}I
\end{bmatrix}
\end{equation}
with $I$ be the identity map and 
\begin{equation}
\hat{b}_{ik}:=\left\{ \begin{array}{lcc}
-1 &  \text{if node $i$ is at the tail of $k$-th edge} 
\\ 1 &  \text{if node $i$ is at the head of $k$-th edge} \\
 0 &  \text{otherwise}.
\end{array}
\right.
\end{equation}
The adjoint operator of $B$ is given by the map   $B^*:\Lambda^0\rightarrow \Lambda^1$    with matrix representation $\hat{B}^{\top}\otimes I$, and it is called the \emph{co-incidence operator}. We will throughout use $B$ ($B^{\top}$) both for the incidence (respectively, co-incidence) matrix and for incidence (respectively, co-incidence) operator. The rank of $B$ is at most $N-1$ due to  the sum of its rows is zero. Indeed, the rank is $N-1$ when the graph is connected. The columns of $B$ are linearly independent. We also introduce the \emph{Laplacian matrix} given by $ {L}:=BB^{\top}$. 

\subsection{Agents and network dynamics}
We consider agents evolving on a configuration manifold $\mathcal{Q}_i$  of dimension $n$ for all $i\in\{1,\dots,N\}$. The position of the $i$-th robot is given by the vector $q_i\in\mathcal{Q}_i$ and velocity $\dot{q}_i\in T_{q_i}\mathcal{Q}_i$. The equation of motion of the $i$-th agent are given  by the Euler-Lagrange equations
\begin{equation}
\begin{split}
\dot{q}_i&=v_i,\\
M_i(q_i)\dot{v}_i+C_i(q_i,v_i)v_i+g_i(q_i)&=\tau_i,    
\end{split}
\label{eq:euleractual}
\end{equation}
where   $\tau_i\in T^*_{q_i}\mathcal{Q}$ are   external forces  $g_i(q_i)=\frac{\partial P_i}{\partial q_i}(q_i)$ and  $C_i(q_i,v_i)$ is any matrix  satisfying
\begin{equation}
\begin{split}
C_i(q_i,v_i)v_i&=\dot{M}_i(q_i)v_i-\frac{\partial}{\partial q_i}\left(\frac{1}{2}v_i^{\top}M_i(q_i)v_i\right).\\
\end{split}
\label{eq:coriolis1}
\end{equation}
When  $C_i(q_i,v_i)$  is expressed in terms of the  so-called Christoffel symbols, it leads to the following   property
\begin{lemma}[\cite{arjanl2}]
	The matrix
	\begin{equation}
	N_i(q_i,v_i)=\dot{M}_i(q_i)-2C_i(q_i,v_i),
	\label{eq:skeqsymmetricEuler}
	\end{equation}
	is linear in $v_i$ and skew-symmetric for every $q_i$ and $v_i$.
\end{lemma} 
This property is a clear expression that the forces $N_i(q_i,v_i)v_i$ are \emph{workless}\footnote{Notice that, also in case that $C_i(q_i,v_i)$ is not parametrized by the Christoffel symbols,  by energy conservation, matrix   $N_i(q_i,v_i)$ still satisfies condition  $v_i^{\top}{N}_i(q_i,v_i)v_i=0$ for every $q_i$ and $v_i$.}. Indeed, if we take the total co-energy  $H^*_i(q_i,v_i)=\frac{1}{2}\dot{q}_i^{\top}M_i(q_i)\dot{q}_i+P_i(q_i)$ as storage function,  then system \eqref{eq:euleractual} is  passive (\emph{lossless}), i.e., $\dot{H}^*_i(q_i,\dot{q}_i)=\dot{q}^{\top}N_i(q_i,\dot{q}_i)\dot{q}+y_i^{\top}\tau_i=y_i^{\top}\tau_i$ where the  output is $y_i=v_i$. 

The EL network dynamics can be expressed in a compact form by using the concatenated vectors,
\begin{equation*}
\begin{split}
q:=[q_1^{\top},\dots,q_N^{\top}]^{\top},\quad v:=[v_1^{\top},\dots,v_N^{\top}]^{\top},\\ \tau:=[\tau_1^{\top},\dots,\tau_N^{\top}]^{\top},\quad g:=[g_1^{\top},\dots,g_N^{\top}]^{\top},
\end{split}
\end{equation*} 
together\footnote{Some function arguments are left out due to space limitations.} with  the block diagonal matrices
\begin{equation*}
\begin{split}
M(q)&:=\text{diag}(M_1(q_1),\dots,M_N(q_N)),\\
C(q,v)&:=\text{diag}(C_1(q_1,v_1),\dots,C_N(q_N,v_N)),
\end{split}
\end{equation*}
such that the dynamics of the network with state $(q,v)$ is given by
\begin{equation}
\begin{split}
\dot{q}&=v,\\
M(q)\dot{v}+C(q,v)v+g(q)&=\tau,
\end{split}
\label{eq:networkdynamics-open1}
\end{equation}
where, as before, $g(q)=\frac{\partial P}{\partial q}(q)$ with the potential function $P(q):=\sum_{i=1}^{N}P_i(q_i)$. Notice that network dynamics \eqref{eq:networkdynamics-open1} preserves the  structure of the $i$-th agent in \eqref{eq:euleractual}, e.g., the matrix $N(q,v):=\dot{M}(q)-2C(q,v)$ is skew-symmetric,  and the map from $\tau$ to $y=v$ is lossless with storage function 
\begin{equation}
H^*(q,v)=\sum_{i=1}^{N}H^*_i(q_i,v_i)=\frac{1}{2}v^{\top}M(q)v+P(q).
\end{equation}

\subsection{Passivity based tracking controller for single agent}

The skew-symmetry of $N_i(q_i,v_i)$ is crucial in the definition of  a \emph{virtual system} associated to \eqref{eq:euleractual}, see e.g.    \cite{arjanl2} and \cite{ortega2013passivity}.
\begin{definition}
	A virtual   system associated to   \eqref{eq:euleractual},  in the state $(r_i,s_i)\in T\mathcal{Q}$,  is defined as  the system
	\begin{equation}
	\begin{split}
	\dot{r}_i&=s_i,\\
	M_i(q_i)\dot{s}_i+C_i(q_i,v_i)s_i+g_i(r_i)&=\overline{u}_i,\\
	y_i&=s_i,  
	\end{split}
	\label{eq:euler-virtual}
	\end{equation}
	parametrized by $(q_i,v_i)$ with input $u_i$ and output $y_i$. 
\end{definition}
Any solution $q_i(t)$ of \eqref{eq:euleractual} for a certain input $\tau_i(t)$  produce   the solution of $(r_i(t),s_i(t))=(q_i(t),\dot{q}_i(t))$ to the virtual system \eqref{eq:euler-virtual} for $\overline{u}_i=\tau_i$, but \emph{not} every pair $(r_i(t),s_i(t))$ is a solution to \eqref{eq:euleractual}. This is only the case if additionally $(r_i(t),s_i(t))=(q_i(t),\dot{q}_i(t))$. Remarkably, system \eqref{eq:euler-virtual} keeps the (structural) lossless property, with the storage function 
\begin{equation}
H^*_i(r_i,s_i)=\frac{1}{2}s^{\top}_iM_i(q_i)s_i+P_i(r_i)
\label{eq:storageVirtualLagrange}
\end{equation}
parametrized by $q_i$. The time derivative a long \eqref{eq:euler-virtual} is
\begin{equation}
\begin{split}
\dot{H^*}(r_i,s_i)&=\frac{1}{2}s_i^{\top}\dot{M}(q)s_i+s_i^{\top}\left(u_i-C_i(q_i,v_i)s_i\right),\\
&=s_i^{\top}u_i+s_i^{\top}N_i(q_i,v_i)s_i,\\
&=y_i^{\top}u_i.
\end{split}
\end{equation}
With the feedback $\overline{u}_i=g_i(r_i)-K_is_i+{u}_i$  in  \eqref{eq:euler-virtual} with $K_i>0$  and $\overline{u}_i$ an external input,   position and velocity dynamics are decoupled. The velocity dynamics results in 
\begin{equation}
\begin{split}
M_i(q_i)\dot{s}_i+C_i(q_i,v_i)s_i+K_is_i&={u}_i,\\
y_i&=s_i  ,
\end{split}
\label{eq:eulervelocity-virtual}
\end{equation}
which is output strictly passive with storage function $S_{s,i}(s_i,q_i)=\frac{1}{2}s_i^{\top}M_i(q_i)s_i$ parametrized by $q_i$. From a  design point of view, this fact makes \eqref{eq:eulervelocity-virtual} a suitable  of target closed-loop   system. 

In fact this is the seminal tracking controller in \cite{slotineli}, where a \emph{sliding manifold} is made invariant and attractive by designing the \emph{sliding (error) variable} such that it preserves the velocity dynamics  structure as in the virtual system \eqref{eq:eulervelocity-virtual}. 
\begin{theorem}
	Consider a smooth reference trajectory  $q_d(t)$ together with the change of variables $s_i=v_i-v_{i,r}$ where   $v_{i,r}=\dot{q}_i-\Pi_i(q_i-q_d)$ is an artificial velocity reference with $\Pi_i=\Pi_i^{\top}>0$. Then, velocity dynamics of   \eqref{eq:euleractual}  in closed-loop with the control law
	\begin{equation}
	\begin{split}	\tau_i=M_i(q_i)\dot{v}_{i,r}+C_i(q_i,\dot{q}_i)v_{i,r}+g_i(q_i)-K_is_i+ {u}_i 
	\end{split}
	\label{eq:Fossen}
	\end{equation}
	is given by virtual system \eqref{eq:eulervelocity-virtual} and $s(t)\rightarrow 0$, $\dot{q}(t)- \dot{q}_d(t)\rightarrow 0$ and  $q(t)-q_d(t) \rightarrow 0$ exponentially as $t\rightarrow \infty$. 
	\label{lemma:SlotineLi}
\end{theorem} 
This control law gives exponential stability as shown in \cite{spong1990}. A \emph{backstepping redesign}  is proposed  in  \cite{fossen1997nonlinear}  where  the closed-loop system performance is further improved as given in the following theorem.
\begin{theorem}
	Consider the change of coordinates $\tilde{q}_i=q_i-q_d$  and all hypothesis and variables of Theorem \ref{lemma:SlotineLi}. Then, for  system \eqref{eq:euleractual}  in closed-loop with the control law
	\begin{equation}
	\begin{split}
\tau_i=M_i(q_i)\dot{v}_{i,r}+C_i(q_i,\dot{q}_i)v_{i,r}+g_i(q_i)-K_is_i-\Pi_i\tilde{q}_i+{u}_i 
	\end{split}
	\label{controllawlemma3}
	\end{equation}
the origin  $(\tilde{q}_i,s_i)=(0,0)$ is globally uniformly exponentially  stable equilibrium point.
\end{theorem}

\section{Distributed node \& edge dynamic  controller design}\label{sec:PassivityTool}
\subsection{Group coordination problem formulation}\label{sec:Group-cood-problem}
In this paper, we are interested in the following group coordination problem: For a network of EL systems in \eqref{eq:euleractual}, design a distributed control law on each node and  edge based only on local information\footnote{The $i$-th agent can use the information $y_i-y_j$ if the $j$-th agent is a neighbor, where $y_i$ is the \emph{passive output} of the $i$-th agent in \eqref{eq:euleractual}.} such that  
\begin{itemize}
	\item Each agent's velocity $v_i(t)$ in \eqref{eq:euleractual} tracks a common velocity reference $v_r(t)$ prescribed for the network; that is
	\begin{equation}
	\lim\limits_{t\rightarrow \infty}\|v_i(t)-v_r(t)\|=0 \quad\text{for all}\quad i\in\{1,\dots,N\}.
	\end{equation}
	\item The interaction variable $\zeta_k$ (defined on the edge and denotes the relative displacement between agents $i$ and $j$ interconnected through an \emph{artificial spring} that will be defined shortly below)  \emph{converges to a nonempty compact set} $\mathcal{A}_k\subset \mathcal{Q}$, for all $k\in\{1,\dots,M\}$.
    \end{itemize}

\subsection{Node \& edge dynamic control design method}
Our proposed design procedure is as follows: 
\begin{enumerate}
	\item 
    For each EL agent \eqref{eq:euleractual}, consider the nodal feedback control \eqref{eq:Fossen} such that the closed-loop system is passive from an external input ${u}_i$ to the velocity "error" $y_i:=s_i=v_i-v_r$. Denote this local closed-loop system as $y_i=\mathcal{H}_i\{u_i\}$.
	\item  For every edge $k$, we assign a (spring) system 
	\begin{equation}
	\begin{split}
	\Sigma_{\mu,k}^z:\left\{ \begin{array}{lcc}
	\dot{\zeta}_k=\mu_k-\phi_{\zeta,k}(\zeta_k),\\
	{\tau_k}= \frac{\partial P_{\zeta,k}}{\partial \zeta_k}(\zeta_k), \quad k\in\{1,\dots,M\},
	\end{array}
	\right.
	\label{eq:Proposition-Art-Springs}
	\end{split}
	\end{equation}	
	to the $k$-th edge if $(i,j)\in \mathcal E_k$, where $\phi_{\zeta,k}$ is a potential force to be designed and  $P_{\zeta,k}:\mathcal{D}_k\subseteq\mathcal{Q}_k\rightarrow \mathds{R}_{\geq 0}$ is a $C^2$ convex artificial potential energy  function defined on the open set $\mathcal{D}_k$  with minimum at $\zeta_d$ where $\zeta_d$ is the vector of desired relative displacement between communicating agents. Note that, in this case, we assume that $\mathcal A_k = \{\zeta_d\}$. 
	\item We interconnect all node systems  $y_i=\mathcal{H}_i\{u_i\}$, $i=1,\ldots, N$ and all edge spring system  $\Sigma_{\mu,k}^z$, $k=1,\dots, M$ through the following passivity preserving interconnection 
	\begin{equation}
	\mu_k:=\sum_{\ell=1}^{N}b_{\ell k}y_{\ell}, \quad\quad u_i=-\sum_{k=1}^{M}b_{ik}\tau_k,
	\label{eq:interconnetionlawa}
	\end{equation}
	for all $k\in\{1,\dots,M\},$ and all $i\in\{1,\dots,N\}$, respectively.
\end{enumerate}

The function $P_{\zeta,k}(\zeta_k)$ is designed to render the target sets $\mathcal{A}_k$ invariant and asymptotically stable for $ k\in\{1,\dots,M\}$.
Interconnection laws in \eqref{eq:interconnetionlawa} satisfy the vector relations
\begin{equation}
\begin{split}
\mu&=(B^{{\top}}\otimes I_n)y \quad\text{and}\quad u= -(B\otimes I_n)\tau,
\end{split}
\label{eq:interconnetionlawa-compact}
\end{equation}
with $y=[y_1^{\top},\dots,y_N^{\top}]^{\top},$ and $ \mu=[\mu_1^{\top},\dots,\mu_M^{\top}]^{\top}$. This means that $\mu$ must be in the image space $\text{Im}(B^{\top}\otimes I_n)$. 
\begin{remark}
We make the following observations with respect to the design procedure in \cite{Arcak}.	
\begin{itemize}
	\item 	If  $\phi_{\zeta,k}(\zeta)=0$ for all $k\in\{1,\dots,M\}$ and $y=v$ in \eqref{eq:interconnetionlawa-compact} then we recover the  procedure in \cite{Arcak}.
	\item  Due to relations in \eqref{eq:interconnetionlawa-compact}, the "symmetric" interconnection structure is preserved.
\end{itemize}
\end{remark}



Similar to the first step in the design procedure in \cite{Arcak}, we also impose that the system $y_i=\mathcal{H}_i\{u_i\}$ to be  \emph{strictly passive} from $u_i$ to $y_i$ with a $C^1$, positive definite, radially unbounded storage function
\begin{equation}
S_i(r_i,s_i)=\frac{1}{2}s^{\top}_iM_i(q_i)s_i+\frac{1}{2}\tilde{q}_i^{\top}\Pi_i\tilde{q}_i,
\label{eq:storageVirtualLagrange-Fossen}
\end{equation}
satisfying 
\begin{equation}
\dot{S}_i(q_i,s_i)\leq W_{y,i}(s_i)+y_i^{\top}u_i,
\end{equation}
for the positive definite function $W_{y,i}(s)=s_i^{\top}K s_i$.

Likewise, for the spring system we also require that its dynamics  \eqref{eq:Proposition-Art-Springs} to be strictly passive. To this end,  take    $\phi_{\zeta,k}(\zeta_k)=\tau_k$. Indeed, with $P_{\zeta,k}(\zeta_k)$ as storage function, the map $\mu_k\mapsto \tau_k$ is strictly passive. This is easily seen from the time derivative of $P_{\zeta,k}(\zeta_k)$ along system  \eqref{eq:Proposition-Art-Springs},
$$\dot{P}_{\zeta,k}=\frac{\partial P_{\zeta,k}^{\top}}{\partial \zeta_k}(\zeta)\mu_k-\underbrace{\frac{\partial P_{\zeta,k}^{\top}}{\partial \zeta_k}(\zeta)\frac{\partial P_{\zeta,k}}{\partial \zeta_k}(\zeta)}_{W_{\zeta,k}(\zeta_k)}.$$
In particular, in this preliminary work we consider 
\begin{equation}
P_{\zeta,k}(\zeta_k)=\frac{1}{2}(\zeta_k-\zeta_d)^{\top}K_{\zeta}(\zeta_k-\zeta_d).
\label{eq:Potential-Spring-k}
\end{equation}
\subsection{Interconnected system stability analysis}

\begin{proposition}
	Consider agent's dynamics  \eqref{eq:euleractual} which is in closed-loop with the control law \eqref{controllawlemma3} 
	for all $i\in\{1,\dots,N\}$, combined with   \eqref{eq:Proposition-Art-Springs} and \eqref{eq:interconnetionlawa-compact}. Then the equilibrium point $(\tilde{q},s,\zeta)=(0,0,\zeta_d)$ is uniformly exponentially stable with rate $\beta=k_3/k_2$, where
	\begin{equation}
	\begin{split}
	k_2&=\max\{\lambda_{\max}(\Pi),\lambda_{\max}(M(q)),\lambda_{\max}(K_{\zeta})\},\\
	k_3&=\min\{\lambda_{min}(\Pi^2),\lambda_{\min}(K),\lambda_{\min}(K_{\zeta}^2)\}.
	\end{split}
	\end{equation}
\end{proposition}
\begin{pf}
	First notice that the closed-loop system is 
	\begin{equation}
	\begin{split}
	\dot{\tilde{q}}+\Pi\tilde{q}&=s,\\
	M(q)\dot{s}+C(q,\dot{q})s+Ks&=-\Pi\tilde{q}-(B\otimes I_n)\frac{\partial P_{\zeta}}{\partial \zeta}(\zeta),\\
	\dot{\zeta}+\frac{\partial P_{\zeta}}{\partial \zeta}(\zeta)&=(B^{\top}\otimes I_n)s
	\end{split}
	\label{eq:closed-loop-Art-Spring}
	\end{equation}
	where in this case   $\Pi:=\text{diag}\{\Pi_1,\dots,\Pi_N\}$ and $K=\text{diag}\{K_1,\dots,K_N\}$. Take as a candidate Lyapunov function 
	\begin{equation}
	S_{\tilde{q}s\zeta}(\tilde{q},s,\zeta,q)=\frac{1}{2}\begin{bmatrix}
	\tilde{q}\\
	s
	\end{bmatrix}^{\top}\begin{bmatrix}
	\Pi & 0 \\
	0 & M(q) \\
	\end{bmatrix}\begin{bmatrix}
	\tilde{q}\\
	s
	\end{bmatrix}+P_{\zeta}(\zeta),
	\label{eq:Lyapunov-spring}
	\end{equation}
	which satisfies the bounds
	\begin{equation}
	k_1\bigg\| \begin{bmatrix}
	\tilde{q}\\
	s\\
	\zeta-\zeta_d
	\end{bmatrix} \bigg\|^2\leq S_{\tilde{q} s \zeta}(\tilde{q},s,\zeta,q) \leq k_2\bigg\| \begin{bmatrix}
	\tilde{q}\\
	s\\
	\zeta-\zeta_d
	\end{bmatrix} \bigg\|^2
	\label{eq:eq:storagefunction-slotinelie-network-Springs-bounds}
	\end{equation}
	where 
	\begin{equation*}
	k_1=\min\{\lambda_{min}(\Pi),\lambda_{min}(M(q)),\lambda_{min}(K_{\zeta})\}.
	\end{equation*}
	The time derivative of \eqref{eq:Lyapunov-spring} along   \eqref{eq:closed-loop-Art-Spring} is given by 
	\begin{equation}
	\dot{S}_{\tilde{q}s\zeta}=-\tilde{q}^{\top}\Pi^2\tilde{q}-s^{\top}Ks-(\zeta-\zeta_d)^{\top}K_{\zeta}^{2}(\zeta-\zeta_d)<0.
	\end{equation}
	That is, \eqref{eq:Lyapunov-spring} is a strict Lyapunov function for system \eqref{eq:closed-loop-Art-Spring}. Furthermore, with the bounds in \eqref{eq:eq:storagefunction-slotinelie-network-Springs-bounds} it is easy to see that $\dot{S}_{\tilde{q}s\zeta}(\tilde{q},s,\zeta,q)<- \beta S_{\tilde{q}s\zeta}(\tilde{q},s,\zeta,q)$ and the function \eqref{eq:Lyapunov-spring} is radially unbounded. This completes the proof. \hfill $\blacksquare$
\end{pf}

\section{Passivity-based synchronized tracking controls}
In this section we present two alternative distributed trajectory tracking control laws which can be seen as particular cases of the method described in Section \ref{sec:PassivityTool} and that also fit with approach in \cite{Arcak} where virtual systems structure is exploited. 
\subsection{Slotine-Li synchronized tracking control}
Recall that the networked EL dynamics \eqref{eq:networkdynamics-open1}  has the same structure as the individual agent dynamics as in \eqref{eq:euleractual}. 
This motivates us to introduce a similar controller construction as the one in Theorem \ref{lemma:SlotineLi} for a single agent. However, in this case,  the control objective is not only tracking to a   reference signal $q_d(t)\in\mathcal{Q}$ but synchronized tracking to a common reference signal $q_d(t)$ for all the agents in the network. To this end,  we propose the following sliding manifold for system \eqref{eq:networkdynamics-open1} given by
\begin{equation}
\Omega=\{(q,v) : \dot{\tilde{q}}+(\Pi\otimes I_n)\tilde{q}+ (B\Delta B^{\top}\otimes I_n){q}=0 \}
\label{eq:network-slidingmanifold}
\end{equation}
where $\tilde{q}=q-(\mathds{1}_N\otimes q_d(t))$  with $\mathds{1}_N\in\mathds{R}^N$ the vector of all ones,  $\Pi=\Pi^{\top}\in\mathds{R}^{N\times N}$ and $\Delta=\Delta^{\top}\in\mathds{R}^{M\times M}$ are positive definite diagonal  matrices. Since $ (B\Delta B^{\top}\otimes I_n)(\mathds{1}_N\otimes q_d(t))=0_N$, the  dynamics of \eqref{eq:networkdynamics-open1} in the manifold $\Omega$ in \eqref{eq:network-slidingmanifold} satisfies
\begin{equation}
\dot{\tilde{q}}=-([\Pi+B\Delta B^{\top}]\otimes I_n)\tilde{q}.
\label{eq:network-slidingdynamids}
\end{equation}
Thus if  $\Pi$ and $\Delta$ are chosen such that they satisfy 
\begin{equation}
P([\Pi+B\Delta B^{\top}]\otimes I_n )+([\Pi+B\Delta B^{\top}]\otimes I_n)^{\top}P=-Q
\end{equation}
where $P, Q \in\mathds{R}^{Nn\times Nn}$ are symmetric and positive definite matrices, then $q(t)\rightarrow( \mathds{1}_N\otimes q_d(t))$ exponentially as $t \rightarrow \infty$. Hence all the agents track $q_d(t)$ in a synchronized fashion. 
Thus, by defining 
\begin{equation}
v_r:=(\mathds{1}\otimes \dot{q}_d(t))-([\Pi+ B\Delta B^{\top}] \otimes I_n)\tilde{q},
\label{eq:artificial-virtual-reference-velocity}
\end{equation}
as an artificial velocity reference signal for the network dynamics \eqref{eq:networkdynamics-open1}, the sliding variable is given by $s=\dot{q}-v_r$.
\begin{corollary}\label{corollary:sync-slotine-li}
	Consider a  strongly connected graph $\mathcal{G}$ with incidence matrix $B$ and a reference trajectory $(\mathds{1}_N\otimes q_d(t))$ for system \eqref{eq:networkdynamics-open1}  with $q_d(t)\in\mathcal{Q}$. Let $s$, $v_r$ be the sliding variable and artificial velocity reference signal as defined before and the control law be given by
	\begin{equation}
	\begin{split}
\tau=M(q)\dot{v}_{r}+C(q,\dot{q})v_{r}+g(q)-Ks+{u}
	\label{eq:slotine-li-sync-tracking}
	\end{split}
	\end{equation}	
	where  $K=K^{\top}\in\mathds{R}^{Nn\times Nn}$ is a positive definite gain matrix. Then the closed-loop system of \eqref{eq:networkdynamics-open1} and \eqref{eq:slotine-li-sync-tracking}  
	\begin{equation}
	\begin{split}
	\dot{\tilde{q}}+([\Pi+B\Delta B^{\top}]\otimes I_n )\tilde{q}&=s,\\
	M(q)\dot{s}+C(q,v)s+Ks&={u},
	\label{eq:networkdynamics-closedloop}
	\end{split}
	\end{equation}
	defines a strictly passive map $u \mapsto y=s$ with respect to the storage function
	\begin{equation}
	S_s(s,q)=\frac{1}{2}s^{\top}M(q)s,
	\label{eq:storagefunction-slotinelie-network}
	\end{equation}
	parametrized by $q$. Moreover, by taking $u=\alpha(s)$ with a passive map $\alpha$, we have that $s(t)\rightarrow 0$ and $\tilde{q}(t)\rightarrow 0$ as $t\rightarrow \infty$.
\end{corollary}

Corollary \ref{corollary:sync-slotine-li} can be seen as a particular case of the   result in \cite[Theorem 6.3]{bai2011cooperative}. The only difference is that  we consider    the (artificial) reference velocity $v_r$ defined above (that is not common to all the agents),  instead of a pure time-varying signal. The structure of $v_r$ in this case implies that the external contains also a diffusive velocity coupling, i.e, 
\begin{equation}
\overline{u}=-(B\Delta B^{\top}\otimes I_n)(q+v)+{u}.
\end{equation} 
Nevertheless, Corollary \ref{corollary:sync-slotine-li} can 
be proved exactly in the same way as in  \cite[Theorem 6.3]{bai2011cooperative}. The existence of   $\Omega$ in \eqref{eq:network-slidingmanifold} is guaranteed with $v_r$ as defined above. This in turn implies that   both $(\mathds{1}_N\otimes q_d(t))$ and  $v_r$ are attractive trajectories for \eqref{eq:networkdynamics-closedloop-fossen}. Furthermore, the control scheme \eqref{eq:slotine-li-sync-tracking} can be split into  the so-called \emph{equivalent control}\footnote{The term equivalent control is adopted in the sliding mode control literature \cite{utkin}. This scheme can also be interpreted as an on-manifold control of the I$\&$I framework  in  \cite{astolfi2003immersion}.} with  $\tau_{eq}=M(q)\dot{v}_{r}+C(q,\dot{q})v_{r}+g(q)$ and a feedback term $\tau_{fb}=-Ks$; the first ensures invariance once constrained to $\Omega$   and the later ensures  that the off-manifold "distance" $s$ converges to zero, i.e.,  attractivity. The sliding dynamics is given in \eqref{eq:network-slidingdynamids}, modulus a change of coordinates.
\subsection{Backstepping synchronized tracking control}
In the previous distributed control approach, passivity of  the closed-loop dynamics \eqref{eq:networkdynamics-closedloop} as a whole is not used. This can be further exploited for performance improvement. To this end,  notice that the position error dynamics is passive from  the input  $\overline{u}_{q}=s$ to  the output $y_q=\tilde{q}$  with  the storage function 
\begin{equation}
S_{\tilde{q}}(\tilde{q})=\frac{1}{2}\tilde{q}^{\top}([\Pi+B\Delta B^{\top}]\otimes I_n )\tilde{q}.
\end{equation}
It is \emph{hierarchically} interconnected to  the  passive dynamics of $s$ in \eqref{eq:networkdynamics-closedloop} via $\overline{u}_q=s$, that is, $\tilde{q}$-dynamics does not influence $s$-dynamics. The above observations motivate the following proposition where we  apply a backstepping  {redesign}   for the protocol in Corollary \ref{corollary:sync-slotine-li}, which can be seen as the networked version of  \eqref{controllawlemma3}.

\begin{proposition}
	Consider a  strongly connected graph $\mathcal{G}$ with incidence matrix $B$, and a reference trajectory $(\mathds{1}_N\otimes q_d(t))$ for system \eqref{eq:networkdynamics-open1}  with $q_d(t)\in\mathcal{Q}$,  $s=v-v_r$,  and  $v_r$ as in \eqref{eq:artificial-virtual-reference-velocity} together with the control law given by
	\begin{equation}
	\begin{split}
	\tau=M(q)\dot{v}_{r}+C&(q,\dot{q})v_{r}+g(q)-Ks\\&-([\Pi+B\Delta B^{\top}]\otimes I_n )\tilde{q}+{u}
	\label{eq:slotine-li-sync-tracking-fossen}
	\end{split}
	\end{equation}	
	where  $K=K^{\top}\in\mathds{R}^{Nn\times Nn}$ is a positive definite gain matrix.  Then the closed-loop system of \eqref{eq:networkdynamics-open1} with the control law \eqref{eq:slotine-li-sync-tracking-fossen} given by 
	\begin{equation}
	\begin{split}
	\dot{\tilde{q}}+([\Pi+B\Delta B^{\top}]\otimes I_n )\tilde{q}&=s,\\
	M(q)\dot{s}+C(q,\dot{q})s+Ks+([\Pi+B\Delta B^{\top}]\otimes I_n )\tilde{q}&={u},
	\label{eq:networkdynamics-closedloop-fossen}
	\end{split}
	\end{equation}
	is strictly passive from ${u}$ to $s$ with the storage function
	\begin{equation}
	S_{\tilde{q}s}(\tilde{q},s,q)=S_{\tilde{q}}(\tilde{q})+S_{s}(s,q)
	\label{eq:storagefunction-slotinelie-network-fossen}
	\end{equation}
	parametrized by $q$. Moreover the origin of \eqref{eq:networkdynamics-closedloop-fossen} is uniformly globally exponentially stable with the rate $\beta=k_3/k_2$ where
	\begin{equation}
	\begin{split}
	k_2&=\max\{\lambda_{\max}([\Pi+B\Delta B^{\top}]\otimes I_n),\lambda_{\max}(M(q))\},\\	
	k_3&=\min\{\lambda_{\min}([\Pi+B\Delta B^{\top}]^2\otimes I_n), \lambda_{\min}(K)\}.
	\label{eq:boundsFossen}
	\end{split}
	\end{equation} 
	with $\lambda_{\min}(\cdot)$ and $\lambda_{\max}(\cdot)$ are the minimum and maximum eigenvalue of their argument, respectively.
\end{proposition}
\begin{pf}
	We will show that \eqref{eq:storagefunction-slotinelie-network-fossen} is a \emph{strict Lyapunov function} for the  nonautonomous system \eqref{eq:networkdynamics-closedloop-fossen} following  \cite[Theorem 4.10]{khalil2002noninear}. First, we notice that   \eqref{eq:storagefunction-slotinelie-network-fossen} satisfies
	\begin{equation}
	k_1\bigg\| \begin{bmatrix}
	\tilde{q}\\
	s
	\end{bmatrix} \bigg\|^2\leq S_{\tilde{q} s}(\tilde{q},s,q) \leq k_2\bigg\| \begin{bmatrix}
	\tilde{q}\\
	s
	\end{bmatrix} \bigg\|^2
	\label{eq:eq:storagefunction-slotinelie-network-fossen-bounds}
	\end{equation}
	where 
	\begin{equation*}
	k_1=\min\{\lambda_{min}([\Pi+B\Delta B^{\top}]\otimes I_n),\lambda_{min}(M(q))\},
	\end{equation*}
	and $k_2$ is as in \eqref{eq:boundsFossen}. The time derivative of \eqref{eq:storagefunction-slotinelie-network-fossen} along the trajectories of \eqref{eq:networkdynamics-closedloop-fossen} is
	\begin{equation}
	\dot{S}_{\tilde{q} s}(\tilde{q},s,q)=-\begin{bmatrix}
	\tilde{q}\\
	s
	\end{bmatrix}^{\top}\begin{bmatrix}
	([\Pi+B\Delta B^{\top}]^2\otimes I_n) & 0\\
	0 & K
	\end{bmatrix}\begin{bmatrix}
	\tilde{q}\\
	s
	\end{bmatrix}<0,
	\end{equation}
	uniformly in $q$. Hence, the storage function \eqref{eq:storagefunction-slotinelie-network-fossen} qualifies as a strict Lyapunov function. Using the bounds of the Lyapunov function in \eqref{eq:eq:storagefunction-slotinelie-network-fossen-bounds}, it is straightforward to see that     $\dot{S}_{\tilde{q} s}(\tilde{q},s,q)\leq -\beta {S}_{\tilde{q} s}(\tilde{q},s,q)$, which completes the proof. \hfill $\blacksquare$
\end{pf}
The extra   term in the protocol \eqref{eq:slotine-li-sync-tracking-fossen} can be understood as a \emph{feedforward action} to the closed-loop  dynamics of the sliding variable $s$ that preserves the passivity.

\begin{remark}
	Since system \eqref{eq:networkdynamics-closedloop-fossen} is linear in the state $(\tilde{q},s)$, invoking \emph{contraction analysis} arguments, it can be shown that the matrix
	\begin{equation}
	\begin{bmatrix}
	([\Pi+B\Delta B^{\top}]\otimes I_n) & 0\\
	0 & M(q)
	\end{bmatrix}
	\end{equation}
	is a \emph{Riemannian metric}, for   details see \cite{jouffroy} and references therein. Thus, the  gain matrices $\Pi,\Delta$ and $K$  can be optimally tuned by taking  \eqref{eq:storagefunction-slotinelie-network-fossen}  as a cost function subject   to the network dynamics \eqref{eq:networkdynamics-closedloop-fossen}. 
\end{remark}

\section{Simulations}
To  show the performance of the coordination protocol obtained in Section \ref{sec:PassivityTool}, consider a network of one degree of freedom systems ($n=1$)  with $N=6$ on an undirected graph $\mathcal{G}$ with the following  specifications:
\begin{table}[h!]
\centering
\caption{Simulation parameters}
\label{table:simulation}
\begin{tabular}{|l|l||l|l|}\hline\hline
$q_d=$ & 0.36 & $\Pi_i=$ & 3.5  \\\hline
$K_i=$ & 12 & $K=$ & 5  \\\hline
$z_d=$ & 0 &  &   \\\hline
\end{tabular}
\end{table}

\begin{figure}[h!]
  \centering
  \includegraphics[width=0.5\textwidth]{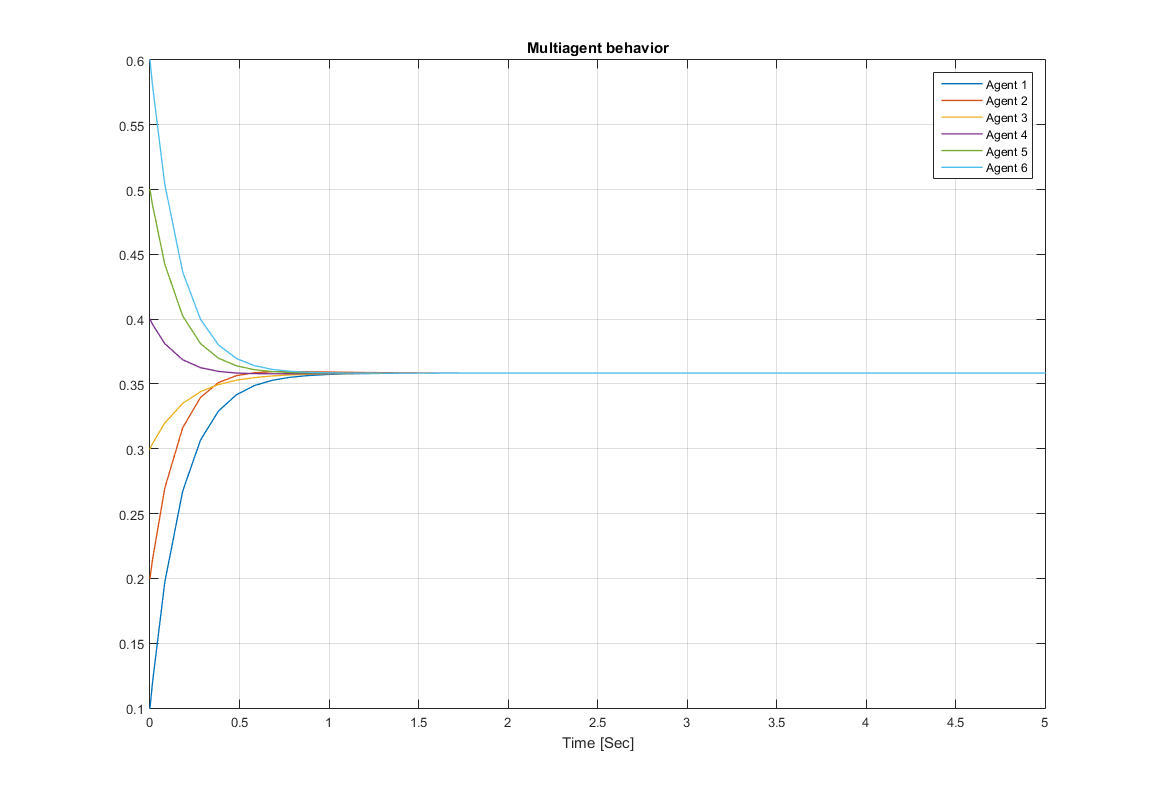}
  \label{fig:simulation}
  \caption{Position $q_i$ vs time for $i\in\{1,2,3,4,5,6\}.$}
  \end{figure}

Due to space limitations, we show only the position tracking behavior in Figure \ref{fig:simulation}, where in fact, the positions reach agreement for the given conditions. 
  
\section{Conclusions}
In this paper we have reformulated the design procedure in \cite{Arcak}, where an artificial spring system is designed at each edge in the graph, instead of diffusing information through the relative positions and velocities, as commonly adopted. 
In our proposed approach, we require that all the nodes' and edges' dynamics to be strictly passive such that via a  passivity preserving interconnection, the total storage function can be used as a strict Lyapunov function to show exponential converge to the desired trajectory. 
\bibstyle{ifacconf}
\bibliography{bibliografia}             

\begin{thebibliography}{23}
\providecommand{\natexlab}[1]{#1}
\providecommand{\url}[1]{\texttt{#1}}
\providecommand{\urlprefix}{URL }
\expandafter\ifx\csname urlstyle\endcsname\relax
  \providecommand{\doi}[1]{doi:\discretionary{}{}{}#1}\else
  \providecommand{\doi}{doi:\discretionary{}{}{}\begingroup
  \urlstyle{rm}\Url}\fi

\bibitem[{Andrieu et~al.(2016)Andrieu, Jayawardhana, and
  Praly}]{andrieu2016tac}
Andrieu, V., Jayawardhana, B., and Praly, L. (2016).
\newblock Transverse exponential stability and applications.
\newblock \emph{IEEE Transactions on Automatic Control}, 61(11), 3396--3411.

\bibitem[{Arcak(2007)}]{Arcak}
Arcak, M. (2007).
\newblock Passivity as a design tool for group coordination.
\newblock \emph{IEEE Transactions on Automatic Control}, 52(8), 1380--1390.

\bibitem[{Astolfi and Ortega(2003)}]{astolfi2003immersion}
Astolfi, A. and Ortega, R. (2003).
\newblock Immersion and invariance: A new tool for stabilization and adaptive
  control of nonlinear systems.
\newblock \emph{IEEE Transactions on Automatic control}, 48(4), 590--606.

\bibitem[{Bai et~al.(2011)Bai, Arcak, and Wen}]{bai2011cooperative}
Bai, H., Arcak, M., and Wen, J. (2011).
\newblock \emph{Cooperative control design: a systematic, passivity-based
  approach}.
\newblock Springer Science \& Business Media.

\bibitem[{Bollob{\'a}s(1998)}]{bollobas1998modern}
Bollob{\'a}s, B. (1998).
\newblock \emph{Modern graph theory (Graduate texts in mathematics)}.
\newblock Springer New York.

\bibitem[{Chopra and Spong(2006)}]{chopra2006passivity}
Chopra, N. and Spong, M.W. (2006).
\newblock Passivity-based control of multi-agent systems.
\newblock In \emph{Advances in robot control}, 107--134. Springer.

\bibitem[{Chung and Slotine(2009)}]{Chung}
Chung, S.J. and Slotine, J.J.E. (2009).
\newblock Cooperative robot control and concurrent synchronization of
  lagrangian systems.
\newblock \emph{IEEE Transactions on Robotics}, 25(3), 686--700.
\newblock \doi{10.1109/TRO.2009.2014125}.

\bibitem[{Fossen and Berge(1997)}]{fossen1997nonlinear}
Fossen, T.I. and Berge, S.P. (1997).
\newblock Nonlinear vectorial backstepping design for global exponential
  tracking of marine vessels in the presence of actuator dynamics.
\newblock In \emph{Decision and Control, 1997., Proceedings of the 36th IEEE
  Conference on}, volume~5, 4237--4242. IEEE.

\bibitem[{Garcia~de Marina~Peinado et~al.(2018)Garcia~de Marina~Peinado,
  Jayawardhana, and Cao}]{marina2018}
Garcia~de Marina~Peinado, H., Jayawardhana, B., and Cao, M. (2018).
\newblock Taming inter-distance mismatches in formation-motion control for
  rigid formations of second-order agents.
\newblock \emph{IEEE Transactions on Automatic Control}, 63(2), 449--462.

\bibitem[{Jouffroy and Fossen(2010)}]{jouffroy}
Jouffroy, J. and Fossen, T.I. (2010).
\newblock A tutorial on incremental stability analysis using contraction
  theory.
\newblock \emph{Modeling, Identification and control}, 31(3), 93--106.

\bibitem[{Khalil(2002)}]{khalil2002noninear}
Khalil, H.K. (2002).
\newblock \emph{Noninear systems}, volume~3.
\newblock Prentice-Hall, New Jersey.

\bibitem[{Nu\~no et~al.(2013{\natexlab{a}})Nu\~no, Ortega, Jayawardhana, and
  Basa\~nez}]{nuno2013ejc}
Nu\~no, E., Ortega, R., Jayawardhana, B., and Basa\~nez, L.
  (2013{\natexlab{a}}).
\newblock Networking improves robustness in flexible-joint multi-robot systems
  with only joint position measurements.
\newblock \emph{European Journal of Control}, 19(6), 469--476.

\bibitem[{Nu\~no et~al.(2013{\natexlab{b}})Nu\~no, Ortega, Jayawardhana, and
  Basanez}]{nuno2013automatica}
Nu\~no, E., Ortega, R., Jayawardhana, B., and Basanez, L. (2013{\natexlab{b}}).
\newblock Coordination of multi-agent euler-lagrange systems via
  energy-shaping: Networking improves robustness.
\newblock \emph{Automatica}, 49(10), 3065--3071.

\bibitem[{Ortega et~al.(1998)Ortega, Perez, Nicklasson, and
  Sira-Ramirez}]{ortega2013passivity}
Ortega, R., Perez, J.A.L., Nicklasson, P.J., and Sira-Ramirez, H. (1998).
\newblock \emph{Passivity-based control of Euler-Lagrange systems}.
\newblock Springer Science \& Business Media.

\bibitem[{Ortega and Spong(1988)}]{ortega1988adaptive}
Ortega, R. and Spong, M.W. (1988).
\newblock Adaptive motion control of rigid robots: A tutorial.
\newblock In \emph{Decision and Control, 1988., Proceedings of the 27th IEEE
  Conference on}, 1575--1584. IEEE.

\bibitem[{Reyes~B\'aez et~al.(2017)Reyes~B\'aez, van~der Schaft, and
  Jayawardhana}]{rodolfo2017}
Reyes~B\'aez, R., van~der Schaft, A., and Jayawardhana, B. (2017).
\newblock Tracking control of fully-actuated port-hamiltonian mechanical
  systems via sliding manifolds and contraction analysis.
\newblock In \emph{Proc. 20th World Congress of the International Federation of
  Automatic Control}.

\bibitem[{Reyes-B\'aez et~al.(2018{\natexlab{a}})Reyes-B\'aez, Donaire, van~der
  Schaft, Jayawardhana, and Perez}]{rodolfo2018cdc}
Reyes-B\'aez, R., Donaire, A., van~der Schaft, A., Jayawardhana, B., and Perez,
  T. (2018{\natexlab{a}}).
\newblock Tracking control of marine craft in the port-hamiltonian framework: A
  virtual differential passivity approach.
\newblock In \emph{submitted}.

\bibitem[{Reyes-B\'aez et~al.(2018{\natexlab{b}})Reyes-B\'aez, van~der Schaft,
  and Jayawardhana}]{Lagrangians}
Reyes-B\'aez, R., van~der Schaft, A.J., and Jayawardhana, B.
  (2018{\natexlab{b}}).
\newblock Virtual differential passivity based control for tracking of
  flexible-joints robots.
\newblock In \emph{Proc. 6th IFAC Workshop on Lagrangian and Hamiltonian
  Methods for Nonlinear Control}.

\bibitem[{Slotine and Li(1987)}]{slotineli}
Slotine, J.J.E. and Li, W. (1987).
\newblock On the adaptive control of robot manipulators.
\newblock \emph{The international journal of robotics research}, 6(3), 49--59.

\bibitem[{Spong et~al.(1990)Spong, Ortega, and Kelly}]{spong1990}
Spong, M., Ortega, R., and Kelly, R. (1990).
\newblock Comments on ``adaptive manipulator control: A case study''.
\newblock \emph{IEEE Transactions on Automatic control}, 35(6), 761--762.

\bibitem[{Utkin(2013)}]{utkin}
Utkin, V.I. (2013).
\newblock \emph{Sliding modes in control and optimization}.
\newblock Springer Science \& Business Media.

\bibitem[{Van~der Schaft and Maschke(2013)}]{van2013port}
Van~der Schaft, A. and Maschke, B. (2013).
\newblock Port-hamiltonian systems on graphs.
\newblock \emph{SIAM Journal on Control and Optimization}, 51(2), 906--937.

\bibitem[{Van~der Schaft(2017)}]{arjanl2}
Van~der Schaft, A. (2017).
\newblock \emph{L2-Gain and Passivity Techniques in Nonlinear Control}.
\newblock Springer International Publishing.

\end{thebibliography}
                                                   







\end{document}